\documentclass[]{mandm}
\usepackage[utf8]{inputenc}
\usepackage[T1]{fontenc}
\usepackage{graphicx}
\usepackage{pgfplots} 
\pgfplotsset{compat=1.17}
\usepackage{hyperref}
\usepackage{amsmath}

\begin{document}

\sloppy

\title{Real Time Integration Centre of Mass (riCOM) Reconstruction for 4D-STEM}

\author[CP Yu, T Friedrich et al]{Chu-Ping Yu,$^1$ Thomas Friedrich,$^1$ Daen Jannis,$^1$ ,Sandra Van Aert$^1$ and Johan Verbeeck$^1$}

\affiliation{%
$^1$EMAT, University of Antwerp, Antwerp, Belgium\\
  Corresponding Author: Johan Verbeeck \email{jo.verbeeck@uantwerpen.be}}

\begin{frontmatter}

\maketitle

\begin{abstract}
A real-time image reconstruction method for scanning transmission electron microscopy (STEM) is proposed. With an algorithm requiring only the center of mass (COM) of the diffraction pattern at one probe position at a time, it is able to update the resulting image each time a new probe position is visited without storing any intermediate diffraction patterns. The results show clear features at higher spatial frequency, such as atomic column positions. It is also demonstrated that some common post processing methods, such as band pass filtering, can be directly integrated in the real time processing flow. Compared with other reconstruction methods, the proposed method produces high quality reconstructions with good noise robustness at extremely low memory and computational requirements. An efficient, interactive open source implementation of the concept is further presented, which is compatible with frame-based, as well as event-based camera/file types. This method provides the attractive feature of immediate feedback that microscope operators have become used to, e.g. conventional high angle annular dark field STEM imaging, allowing for rapid decision making and fine tuning to obtain the best possible images for beam sensitive samples at the lowest possible dose.

\vspace{1em}

\end{abstract}
\vspace{-2em}
\end{frontmatter}

\section*{Introduction}
Scanning transmission electron microscopy (STEM) is one of the most powerful tools for inspecting materials with sub-nanometer or even sub-angstrom level resolution. By scanning with a sharp electron probe, information of the sample from each scan position is collected and images that contain features at the atomic level are generated. There are several methods to form images using the data collected from such experiments. Traditionally, detectors that capture electrons from certain ranges of scattering angles are used in the microscope. They generate a value based on the sum of received electrons at each probe position and result in 2D images. Images formed by detectors that collect signals at high scattering angles are even capable of reflecting the scattering power experienced by the electron probe at the corresponding probe position \citep{pennycook1989z}. 

A pixelated detector does not generate a single value, but instead records a convergent beam electron diffraction pattern (CBED) for each probe position by using a large number of pixels, where each pixel can be seen as an individual detector. This results in a 4D dataset (2D CBEDs on a 2D scan grid). More importantly, these advanced direct electron detectors \citep{muller2012scanning, plackett2013merlin, tate2016high} record CBEDs at a much higher rate than traditional charge-coupled device detectors, and allow collecting a regular-sized 4D dataset within a feasible amount of time.

To process 4D datasets, one can define virtual detectors by selecting specific groups of pixels for summation, which result in similar 2D images as traditional detectors would produce, or seek solutions from more advanced and complex methods. Most of these methods take into account the distribution of the electrons on the detector plane, as well as the relationship between CBEDs and their corresponding probe positions, allowing extra information to be extracted from the dataset. %then using traditional methods, and thus image can either be generated with even higher quality or resolution, or that fewer dose are required to form image that contain necessary knowledge for microscopist to analyze their sample. 
This enables reconstructions of higher resolution \citep{nellist1995resolution} and can reduce the dose needed for microscopists to obtain the necessary information to analyse their samples. Within the category of 4D dataset processing methods, iterative optimisation approaches \citep{rodenburg2004phase, maiden2009improved, odstrvcil2018iterative, chen2020mixed, chen2021electron} reconstruct subsets of the full dataset one region at a time. The process repeats and reprocesses each subset until the algorithm converges to a estimated version of electric potential distribution. Other methods that handle 4D datasets without an iterative process, for example single sideband ptychography (SSB) \citep{pennycook2015efficient, yang2015efficient}, centre of mass (COM) or integrated centre of mass (iCOM) \citep{muller2014atomic, yang2015efficient, lazic2016phase, yucelen2018phase} reconstruction methods, have also proven to be much more dose efficient than traditional imaging methods. Compared to iterative process, they are less computationally demanding and guarantee unique solutions since they do not depend on optimisation algorithms. Also, some prior knowledge, such as the prediction of a phase distribution that may arise from astigmatism and defocus, can be provided to this post process for acquiring more detailed information \citep{pelz2021real}. Yet the ability to achieve fast reconstructions, regardless whether they are iterative or not, usually relies on accelerators (e.g. GPU) as well as large amount of computer memory in order to fit in the whole dataset, or some reduced version of it. With an exception of iCOM, most of these post processing methods are thus limited by the hardware to a certain number of probe positions.

Even though these reconstruction methods may be further optimised to reduce the processing time, users still need to wait for the recording of the dataset to be completed before a resulting image can be generated. This waiting time varies, but for datasets composed of a large number of scan points or in situations where the detector has a slow frame rate, this delay would hinder the process of searching for features of interest, as well as adjusting the optical system based on the observations. This process can be achieved by any processing methods that are independent of probe position and do not, or only slightly, take into account the correlation between data, such as traditional imaging methods, COM shift, and some simple derivatives of these signals, such as COM shift divergence \citep{haas2021high}. Although the processing speed of these methods can be very fast due to their simplicity, the amount of electrons needed to generate an image with adequate quality can be too large and increases the risk of destroying the sample before one can even start collecting the data at low dose conditions. As proposed by A. Strauch et al. \citep{strauch2021live}, a live image update of the reconstruction result can be done by first allocating memory for the dataset, and then gradually filling it with collected and processed data during the scanning process. An update of the reconstructed image can be generated anytime by SSB reconstruction, even before the dataset is complete. However, it also indicates that the number of probe positions in a dataset is limited by the GPU memory, as it needs to store data for later processing. At the current state of technology, this approach is  limited in terms of processing rate to about 1000 probe positions per second in the implementation of Strauch et al. \citep{strauch2021live}, while the collection frame rate of direct electron detectors is approaching 100 kHz \citep{pelz2021real} and even the MHz range for event driven cameras at suitable conditions \citep{jannis2021event}.

To overcome these hardware and speed limitations, we hereby propose a new live reconstruction method based on iCOM, which does not rely on storing the entire 4D dataset in memory, does not require accelerators of any kind and thus greatly reduces the computational requirements, as well as allowing reconstructions of images of a larger scale. In this paper, the physical formulation of real-time iCOM (riCOM) is first derived, and details of the software implementation of the reconstruction algorithm are discussed. This software implements a direct interface to the electron camera, and several real-time reconstructed results are recorded, from which one can see that the tuning of the imaging conditions are immediately reflected in live-updated images. RiCOM reconstruction from existing experimental datasets are also shown. These datasets are recorded frame-by-frame or per-event \citep{guo2020electron, jannis2021event}. Both formats can be processed with the riCOM method with little alteration of the algorithm. Reconstruction results with different range of integration and integrated filters are also displayed. They are compared with each other and with other reconstruction methods to put the proposed method into context.%Reconstruction results with different choices of kernel, including different kernel sizes and integrated filters, are also displayed. They are compared with each other and with other reconstruction methods to put the proposed method into context.

\section*{Materials and Methods}
\subsection*{Physical Formulation}

In 4D STEM, the distribution of the electron intensity at each probe position is recorded. The centre of mass of this distribution can then be calculated, resulting in a vector image \(\vec{I}^{COM}(\vec{r}_p)\) or two scalar images describing its x component \(I^{COMx}(\vec{r}_p)\) and y component \(I^{COMy}(\vec{r}_p)\) components. For example,

\begin{equation} \label{eq:1}
I^{COMx}(\vec{r}_p) =  \int\int_{-\infty}^{\infty} k_x I(\vec{k},\vec{r}_p) d^2\vec{k}
\end{equation}

where \(\vec{r}_p\) is the probe position, \(\vec{k}\) indicates a point on the detector plane with components $k_x$ and $k_y$, and \(I\) is the intensity at the pixel at \(\vec{k}\) while the probe is situated at \(\vec{r}_p\). From previous work \citep{lazic2016phase} it follows that (derivation in supplementary)

\begin{equation} \label{eq:2}
\begin{split}
\vec{I}^{COM}(\vec{r}_p) & =  \frac{1}{2}|\psi_{in}(\vec{r}, \vec{r}_p)|^2\star\nabla\phi(\vec{r}) \\
& = \frac{1}{2}\nabla(|\psi_{in}(\vec{r},\vec{r}_p)|^2\star\phi(\vec{r})).
\end{split}
\end{equation}

\begin{figure*}[ht]
\centering
 \includegraphics[width=0.95\textwidth]{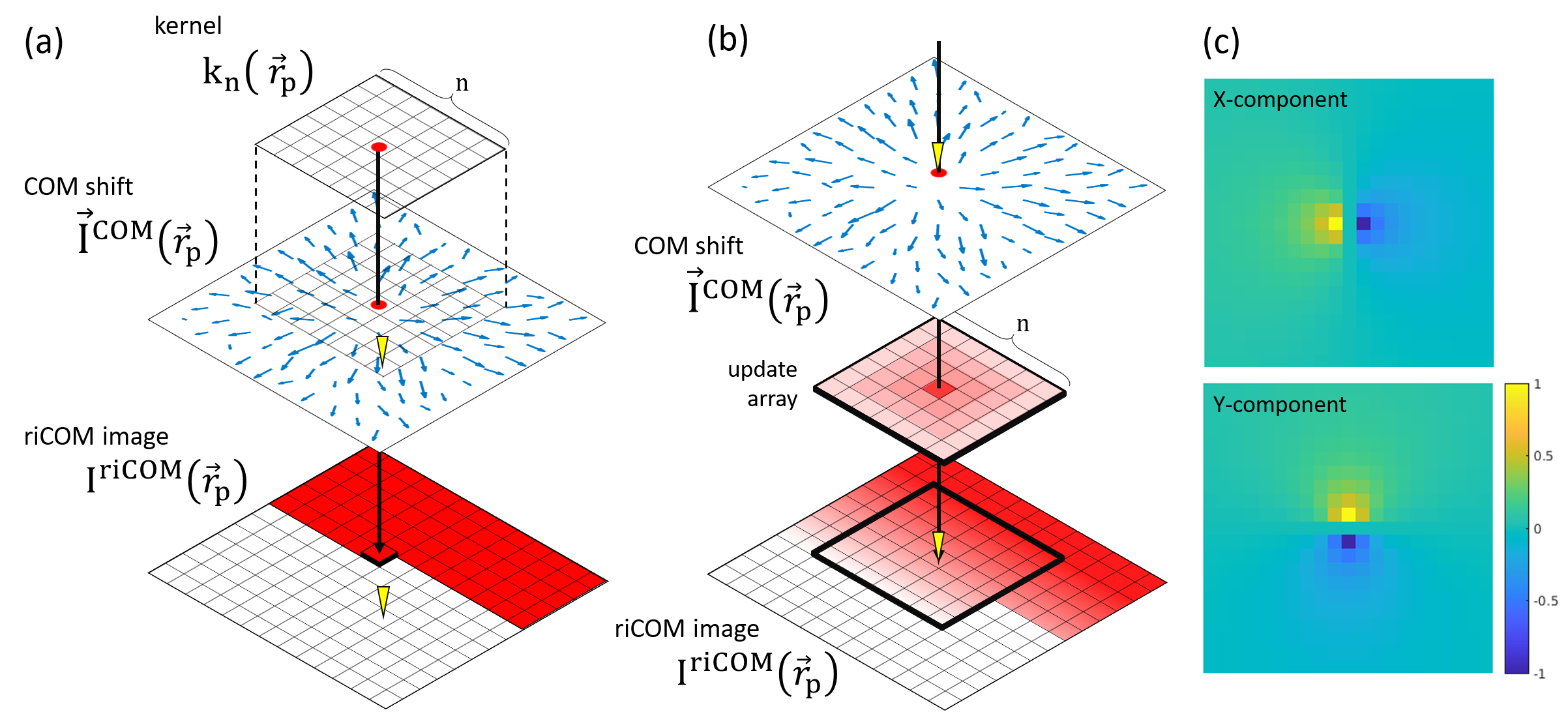}
\caption{
 (a) The kernel takes multiple data points from the COM shift map to calculate the value for one pixel in the riCOM image. (b) The riCOM image is being updated based on the contributions of the COM shift at one probe position. The yellow triangles indicate the scanning probe position. (c) X and Y components of a kernel of size $21\times21$. }
\label{fig_schematic}
\end{figure*}

In equation \ref{eq:2}, the COM shift signal is understood as the gradient (\(\nabla)\) of the local projected potential (\(\phi(\vec{r})\)) cross-correlated with the intensity distribution of the incoming electron beam at a given probe position (\(\psi_{in}(\vec{r},\vec{r}_p)\)).
Note that this result is achieved under the phase object approximation, which assumes that the electron probe remains unmodified while passing through the object. With this approximation the 3D potential established by the material is simplified to a projected potential in a 2D plane. It clearly fails for thicker objects, but it allows a simple derivation and easy understanding of how experimental conditions can affect reconstructed images, which has proven to be a very useful model even for thicker samples in the \hyperref[sec:ResDis]{Result and Discussion section}. 

To solve for a scalar function describing the object, path integration is performed on the COM shift signal to remove the gradient from the right-hand side of equation \ref{eq:2}. For an ideal case the path of the integration can be taken arbitrarily, since the integral is only dependent on the end point of the path integration. However, in realistic cases, the measurement of COM shift contains noise, and thus it would give better estimation of the noise-free result by taking the average of all possible path integrals. By the assumption that equipotential can be found at infinity, this can be achieved by averaging path integrals at all possible azimuthal angles, from infinite afar towards the probe position. In order for this concept to work with a 2D grid of probe positions, the averaged integral can be expressed in a discretised form: 

\begin{equation} \label{eq:3}
\begin{split}
& \frac{1}{2}(|\psi_{in}(\vec{r}, \vec{r}_p)|^2\star\phi(\vec{r})) \\
&= \int_{-\infty}^{\vec{r}_p} \vec{I}^{COM}(\vec{r}) \cdot d\vec{r} \\ 
&= \frac{1}{2\pi} \int_{0}^{2\pi} \int_{-\infty}^{r_p} \vec{I}^{COM}(\vec{r}) \cdot \hat{n} dr d\theta \\ 
&= \frac{a}{2\pi} \sum_{x=-\infty}^{+\infty} \sum_{y=-\infty}^{+\infty} \frac{\vec{r}_p-\vec{r}_{xy}}{|\vec{r}_p-\vec{r}_{xy}|^2} \cdot \vec{I}^{COM}(\vec{r}_{xy}).
\end{split}
\end{equation}

In the continuous representation of the radial averaged path integral (third line of equation \ref{eq:3}), $\hat{n}$ is a unit vector pointing towards $\vec{r}_p$, and in the discrete representation (fourth line of equation \ref{eq:3}), $\vec{r}_{xy}$ describes a vector pointing at each probe position that composes the 2D array.

The discrete representation in equation \ref{eq:3} states that the summation has to go over an infinite amount of points, or at least all probe positions in the dataset (as for iCOM reconstructions) in order to acquire or to approximate the desired object function. This would require the full dataset to be collected first, and rendering a live update of partially reconstructed datasets is therefore impossible. However, it is found that by limiting the spatial range of the summation, the algorithm results in similar reconstructions as iCOM, but with more emphasis on local variations of the object function. This behavior can be understood qualitatively. The term $(\boldsymbol{r_p}-\boldsymbol{r}_{xy}) / |\boldsymbol{r_p}-\boldsymbol{r}_{xy}|^2$ describes an odd function since the vector distribution on both sides of the probe position $\boldsymbol{r_p}$ is the same in magnitude but opposite in direction as the sign changes for $\boldsymbol{r_p}-\boldsymbol{r}_{xy}$. For a global homogeneous COM shift, or for cases where the variation is negligible within the range of the kernel size, it results in an even function for $\boldsymbol{I^{COM}}(\boldsymbol{r}_{xy})$, and thus the product of the two will always be zero. But for short range variations of the object function, which results in local fluctuations of the $\boldsymbol{I^{COM}}$ distribution, it would generate non-zero contribution to the summation result. More detailed discussion of this effect will be covered in the \hyperref[subsec:kerdes]{Kernel Design sub-section}.

\begin{equation} \label{eq:4}
\begin{split}
&I^{riCOM}(\vec{r}_p)  \\
&=\frac{a}{2\pi} \sum_{x=r_{p,x}-\frac{n-1}{2}}^{r_{p,x}+\frac{n-1}{2}}~ \sum_{y=r_{p,y}-\frac{n-1}{2}}^{r_{p,y}+\frac{n-1}{2}} \frac{\vec{r}_p-\vec{r}_{xy}}{|\vec{r}_p-\vec{r}_{xy}|^2} \cdot \vec{I}^{COM}(\vec{r}_{xy})\\
&= k_n(\vec{r}_p) \star \vec{I}^{COM}(\vec{r}_{p})\\
& k_n(\vec{r}_p) = \frac{a}{2\pi} \sum_{x=r_{p,x}-\frac{n-1}{2}}^{r_{p,x}+\frac{n-1}{2}}~ \sum_{y=r_{p,y}-\frac{n-1}{2}}^{r_{p,y}+\frac{n-1}{2}} \frac{\vec{r}_p-\vec{r}_{xy}}{|\vec{r}_p-\vec{r}_{xy}|^2}
\end{split}
\end{equation}

Equation \ref{eq:4} shows the summation with a range $\pm n$ centred at probe position $\vec{r}_p = (r_{p,x}, r_{p,y})$. With this constraint on the range, the integrated centre of mass at one point can be found by only processing COM data from its limited surrounding (Fig. \ref{fig_schematic}-a), allowing data processing to begin and results to be generated during the scanning session. This reconstruction method is thus given the name "real-time iCOM" or "riCOM", as indicated in the same equation by $I^{riCOM}(\vec{r}_p)$. This process is equivalent to a cross-correlation (the $\star$ symbol) between an array $k_n(\vec{r}_p)$ of size $n\times n$, that stores vectors $(\vec{r}_p-\vec{r}_{xy})/|\vec{r}_p-\vec{r}_{xy}|^2$, and the COM shift map $\vec{I}^{COM}(\vec{r}_{xy})$. This array will be referred to as the "kernel" in the rest part of the paper, and images generated by processing COM shift maps with such kernels will be denoted as "riCOM results" or "riCOM images".

Since the kernel processes a group of data points and outputs a value corresponding to the probe position at the centre of the kernel, the collection of data has to lead the reconstruction by $n$ scan lines to fill up the kernel (when scanning in a traditional line by line fashion). This delay between the data collection progress and reconstruction result can be troublesome for operations that highly rely on real time feedback from the scanning process. Since the summation in equation \ref{eq:4} describes a linearly independent process, the contribution from multiple probe positions to a common pixel in the riCOM array can be separately calculated. Furthermore, by collecting contribution from the COM shift at a specific probe position to its vicinity, an update to the riCOM image can be generated in form of an array same size as the kernel. Since this reconstruction scheme depends on one CBED at a time, it leads to an live update of the riCOM result without any delay (Fig. \ref{fig_schematic}-b). Although this does not reduce the time differences between the latest scanning point and the fully updated riCOM pixel, the partially reconstructed fraction of the riCOM image can already show atomic features\footnote{See supplementary documents for example images/videos}, and therefore valuable information at the newly scanned probe position appears with minimal delay. This way the user can also get a quick feedback of their operation. Another advantage is that once the contribution from one probe position is calculated and the corresponding update to the riCOM array is made, the CBED pattern can be discarded, freeing up memory. This effectively removes any memory imposed restriction on scan size if the user is only interested in the resulting riCOM image.

\subsection*{Kernel Design}
\label{subsec:kerdes}

As mentioned in the previous section, summation carried out by a smaller kernel emphasises local object function variations. In other words, it gives more weights to components of higher spatial frequency. To show the relationship between this effect and the kernel size, we start with the Fourier transform of the object function $O(\vec{r}_p)$ for the case of a perfect COM shift measurement (second line of equation \ref{eq:3}). 

\begin{equation} \label{eq:5}
\begin{split}
O(\vec{r}_p)&=\frac{1}{2}(|\psi_{in}(\vec{r}, \vec{r}_p)|^2\star\phi(\vec{r})) \\
 &= \int_{-\infty}^{\vec{r}_p} \vec{I}^{COM}(\vec{r}) \cdot d\vec{r} \\
\mathcal{F} \left\{O(\vec{r}_p)\right\} 
&= \mathcal{F} \left\{\int_{-\infty}^{\vec{r}_p} \vec{I}^{COM}(\vec{r}) \cdot d\vec{r}\right\} \\
&= \mathcal{F} \left\{\vec{I}^{COM}(\vec{r}_p)\right\} \cdot \frac{1}{i\vec{k}_p}.
\end{split}
\end{equation}

Here the symbol $\mathcal{F}$ indicates Fourier transform and $\vec{k}_p$ is a vector in the Fourier domain. As seen in equation \ref{eq:5}, each of the Fourier components of the object function acquires a weight $\frac{1}{i\vec{k}_p}$ after the path integration. This weighting function decays fast with the frequency, and thus low frequency components are attenuated much less then high frequency ones. For riCOM, the result is an averaged integral in the vicinity of the probe position, 

\begin{equation} \label{eq:6}
\begin{split}
& \mathcal{F} \left\{ I^{riCOM}(\vec{r}_p) \right\} \\
&= \mathcal{F} \left\{\int_{\vec{r}_p-\Delta \vec{r}}^{\vec{r}_p} \vec{I}^{COM}(\vec{r}) \cdot d\vec{r} - \int_{\vec{r}_p}^{\vec{r}_p+\Delta\vec{r}} \vec{I}^{COM}(\vec{r}) \cdot d\vec{r}\right\}/2 \\
&= \mathcal{F} \left\{\vec{I}^{COM}(\vec{r}_p)\right\} \cdot \frac{1}{i\vec{k}_p}\times[1-cos(\Delta\vec{r}\cdot\vec{k}_p)] \\
&= \mathcal{F} \left\{O(\vec{r}_p)\right\}\times[1-cos(\Delta\vec{r}\cdot\vec{k}_p)].
\end{split}
\end{equation}

In equation \ref{eq:6}, the riCOM result is approximated by the contribution from both sides of the probe position $\vec{r}_p$ in a single line, within the range of $\Delta\vec{r}$. The result shows that by limiting the integration range, it creates an effect of including another layer of weighting function $1-cos(\Delta\vec{r}\cdot\vec{k}_p)$. This function is flat around zero, and thus strongly suppresses the low frequency signal in the retrieved object function. Also, it peaks at $\vec{k}_p=\frac{\pi}{\Delta\vec{r}}$, which implies that by choosing smaller $\Delta\vec{r}$, or shorter integration range, one can put more weight to the high frequency components. By using kernels with sizes smaller than the real space dimension of the dataset, this effect of limiting integration range can be achieved. Although the actual frequency spectrum of a 2D kernel deviates, the weight of a kernel of size n at each frequency k can be well approximated with the formula derived from line integration

\begin{equation}  \label{eq:freq_approx}
\frac{N}{2k\pi}\times[1-cos(\frac{n-1}{2}\times \frac{2k\pi}{N})].
\end{equation}

Here N is the number of pixels of the image in one direction, and the extra factor $\frac{2\pi}{N}$ scales the pixel size in the Fourier transformed result.

 This effect is not equal to but can be compared to a high-pass filter as it emphasises high frequency details in the reconstructed image. However, other filtering effects, such as low-pass or band-pass filtering, cannot be created simply by altering the kernel size. A filter can be seen as a mask that reduces or eliminates a certain range of frequency signals of an image. This is done by a piece-wise multiplication between the filter and the image in the frequency domain, which is equivalent to a convolution between their real space counterparts. For riCOM images, which can be seen as a cross correlation between a COM shift map and a kernel, the application of such a filter can be included to the design of the kernel.

\begin{equation} 
\label{eq:7}
\begin{split}
I^{riCOM}(\vec{r}_p) * f(\vec{r}_p) &= [\vec{I}^{COM}(\vec{r}_p) \star k_n(\vec{r}_p)] * f(\vec{r}_p) \\
&=\vec{I}^{COM}(\vec{r}_p) \star [k_n(\vec{r}_p) * f(\vec{r}_p)]\\
\end{split}
\end{equation}

\begin{equation} 
\label{eq:8}
\begin{split}
f(\vec{r}_p)) &= \mathcal{F}^{-1} \left\{ F(\vec{k}_p) \right\} \\
F(\vec{k}_p) &= 
\begin{cases}
    1, & k_{max} \geq |k_p| \geq k_{min}\\
    0, & \text{otherwise}
\end{cases}
\end{split}
\end{equation}

In equation \ref{eq:7}, $f(\vec{r}_p)$ is the filter function and the symbol $*$ indicates convolution. Equation \ref{eq:8} writes one of the possible ways to design such a filter, with a hard cutoff at two frequency limits $k_{max}$ and $k_{min}$. The real space counter part of the filter can be found by performing inverse Fourier transform $\mathcal{F}^{-1}$ to the filter function in the Fourier domain. This real space filtering effect can be incorporated to the kernel due the associative property of cross correlation and convolution. It is worth noting that the last part of the equation only holds for central symmetric filters that treat frequency components at different azimuthal angles equally, which is indeed the case for the filter shown in equation \ref{eq:8}. We also want to point out that to create a sharp cutoff at the frequency domain, one would need a filter matching the size of the COM shift array. But in order to keep the size of the kernel, the outcome of the convolution is reduced in size. In other words, the outcome of $k(\vec{r}_p) * f(\vec{r}_p)$ is kept at the same size as $k(\vec{r}_p)$. This would make the cutoff appear in the fashion of a slope and also distorts the rest of the frequency spectrum. 

\begin{figure}[ht]
\centering
\includegraphics[width=0.9\linewidth]{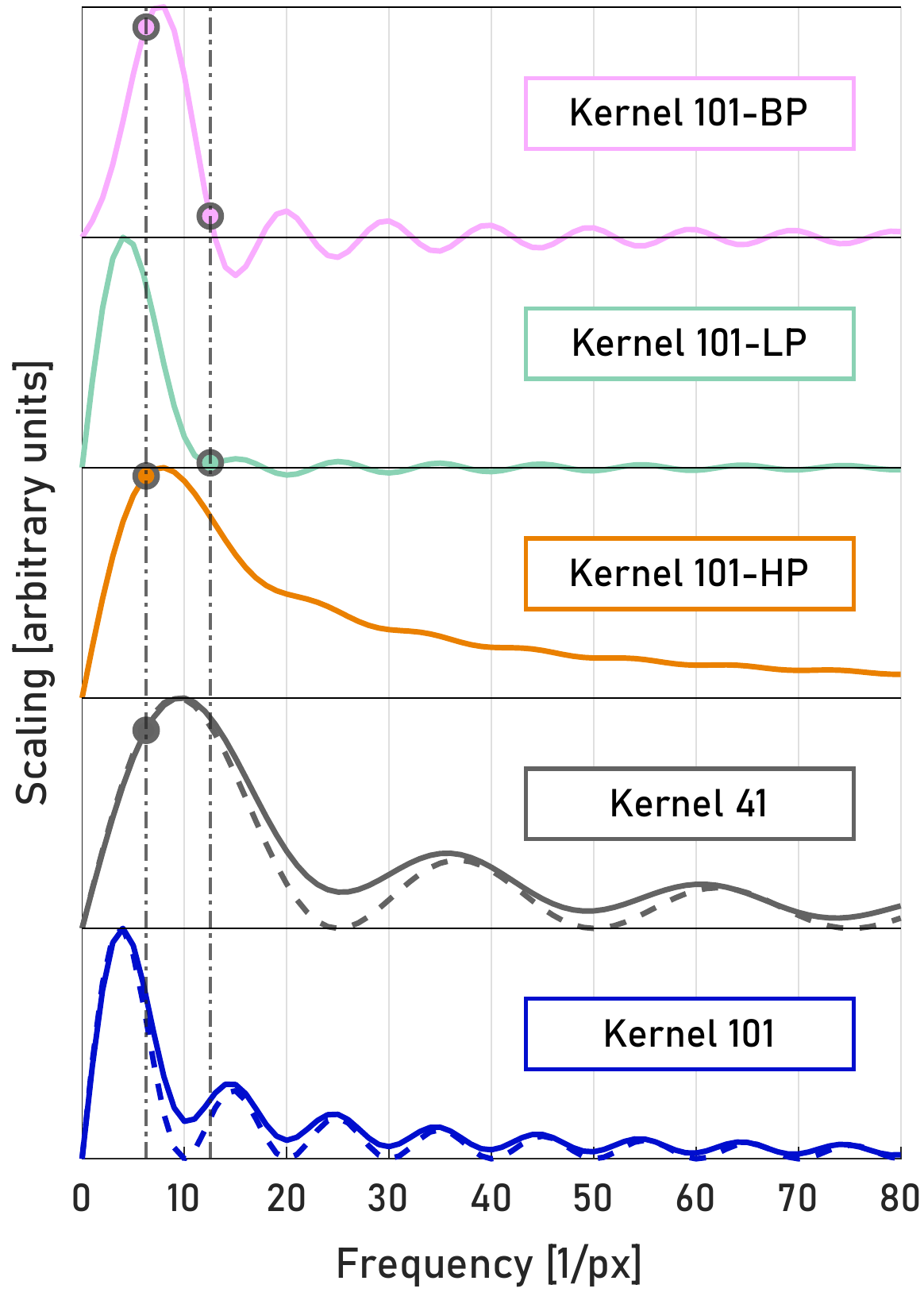}
\caption{
Frequency components of a set of kernels acting on a COM shift map of size $500\times500$. The presented examples include, from bottom to top, the template kernel with size of $101\times101$, a smaller kernel with size of $41\times41$, template kernel with high-pass filter, low-pass filter, and band-pass filter. The dashed line shows the predicted frequency components with line-integration approximation. The two vertical lines indicate the cutoff frequency of the filter or the inverse of the kernel size, and the circles at the intersection of the vertical lines and integral indicate whether a cutoff frequency is applied to the specific design.}
\label{fig_kernel_freq}
\end{figure}

In Fig. \ref{fig_kernel_freq} the frequency components of different kernel designs are illustrated. From bottom to top, the curves corresponds to the template kernel with size of $101\times101$, a smaller kernel with size of $41\times41$, template kernel with high-pass filter ($k_{min} = \frac{12.19}{2}~px^{-1}$), low-pass filter ($k_{max} = 12.19~px^{-1}$), and band-pass filter ($k_{min}, k_{max}$ same as before). For the bottom two curves, the result of the corresponding line integration approximation (dashed lines), with $\vec{\Delta\vec{r}}$ chosen to be half of the kernel size, is also drawn to show their similarity in oscillation frequency and magnitude. 

By comparing the blue and grey curves in Fig. \ref{fig_kernel_freq}, it is clear that Kernel 41 peaks at a higher frequency then Kernel 101, as predicted by the analytical formula, and that the cutoff of the lower frequency due to a smaller kernel happens approximately at the inverse of the kernel size, indicated by the grey circle. This value is then used for $k_{min}$ of the high-pass filter incorporated to Kernel 101-HP (orange curve), which indeed shows a similar overall frequency spectrum as the one of Kernel 41. Notice that the size reduction after convolution between kernel and the decorating filter causes a smooth decrease of frequency components below $k_{min}$, and  spectrum differences beyond $k_{min}$ compared to Kernel 101. 
With the same way, a kernel with low-pass filter \textit{Kernel 101-LP} (green curve) and a kernel incorporating a band-pass filter \textit{Kernel 101-BP}  (pink curve) is created. 
Both of them are showing a suppression of the higher frequency ranges. Kernel 101-BP also shows a shift of the spectrum peak to a higher frequency because of its high-pass characteristic. 

Despite the fact that it is not always possible to recreate the exact characteristics of common post processing filters, the incorporation of filters into the kernel, as well as the choice of kernel sizes allows for a great flexibility for frequency tuning and yields similar and predictable solutions. By combining the kernel and the filter in real space also enables these image processing functions to be applied before the complete riCOM image is rendered, and thus compatible with the live update algorithm.

\subsection*{Data Processing}
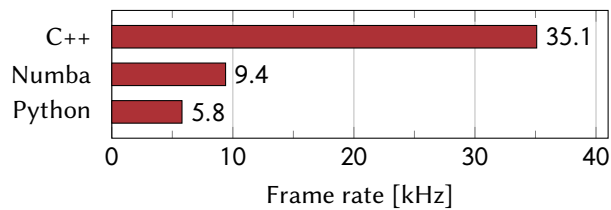
\begin{figure}[htb]
\begin{tikzpicture}
    \begin{axis}[
        xbar,
        y=-0.5cm,
        bar width=0.3cm,
        enlarge y limits={abs=0.35cm},
        width=\linewidth,
        symbolic y coords={C++, Numba, Python},
        ytick={C++, Numba, Python},
        nodes near coords,
        xmin=0,xmax=41,
        xlabel={Frame rate [kHz]},
        xmajorgrids=true,
        minor x tick num={1},
        ytick style={draw=none}
    ]
    \addplot[fill=Maroon] coordinates {(35.1,C++) (9.4,Numba) (5.8,Python)};
\end{axis}    
\end{tikzpicture}
\caption{Speed (Frame Rate in kHz) vs. Implementation benchmark for the computation of riCOM signal with Kernel size of 61x61, data type uint16 and camera size of 256x256 pixels, run on a single thread of an Intel i5-10210U @ 4.2GHz processor. Comparison of a simple implementation in Python, a just-in-time compiled optimisation of the same code, using \href{https://numba.pydata.org/}{Numba} and a version written completely in C++ (compiled with \href{https://gcc.gnu.org/}{GNU gcc-11}).}
\label{fig_comp_time}
\end{figure}

Due to the simplicity of the algorithm, the processing can be carried out completely by CPU with very limited usage of memory. However, in order to reach real time reconstruction that is limited only by the  frame rate of the camera, an efficient implementation of the algorithm is crucial. The benchmark shown in figure \ref{fig_comp_time} shows that an optimised implementation using C++ can easily achieve the maximum speed of $\approx$14~kHz of a MerlinEM camera. Additionally, the pre-processing of binary live data benefits from the low-level features of C++ (e.g. adapting endianess and efficient conversion of binary into numerical arrays). An implementation of the algorithm tailored to event-driven cameras and their corresponding sparse datasets, is even significantly faster. Depending on the dose, >100~kHz have been obtained. The live visualisations at such rates also benefit from using C++ through the possibility of directly accessing and modifying OpenGL textures across threads.

\begin{figure}[ht]
\centering
\includegraphics[width=\linewidth]{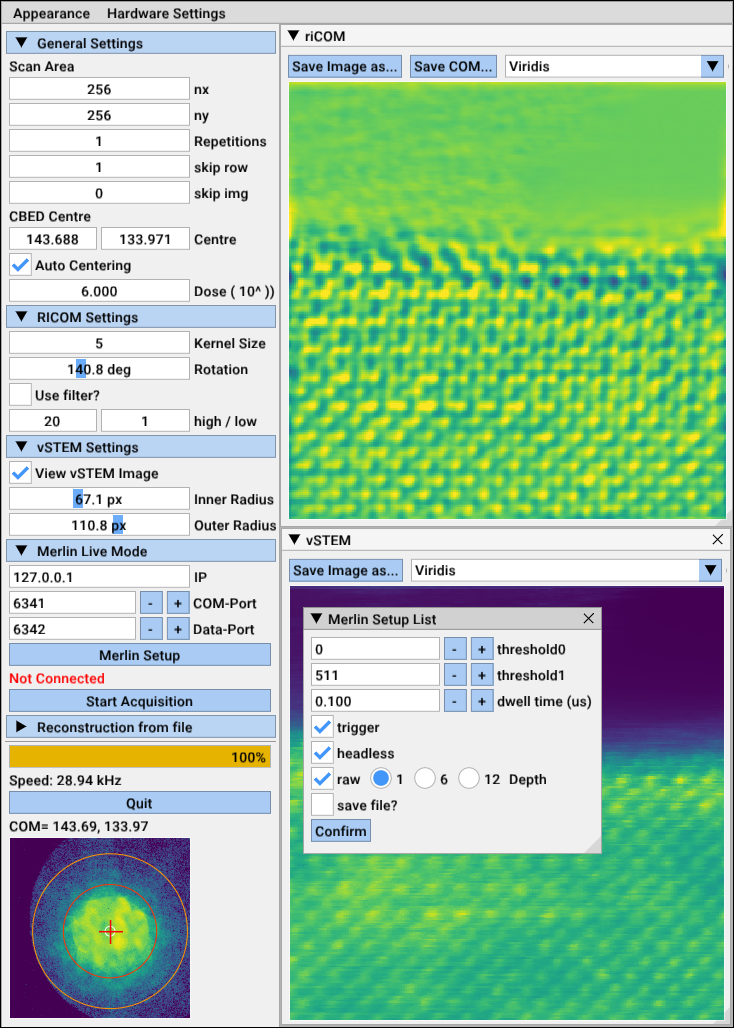}
\caption{Layout of the GUI. The Menu column on the left allows to change various settings, such as scan size, riCOM Kernel and filter settings, virtual STEM settings and interfaces for live mode and file dialogues. During a running reconstruction a CBED is plotted at the bottom of this menu to visually assist interactive tuning of pattern centre and integration area for vSTEM. All other windows are floating panels and can be moved and resized.}
\label{fig_img_gui}
\end{figure}

The program was developed as a cross-platform application that can be run through a command-line-interface (CLI) or interactively through a graphical user interface (GUI) as shown in figure \ref{fig_img_gui}. The core functionality of the algorithm is implemented in a single C++ class object. Visual interfaces interact with an instance of that class across threads through pointers, which allows live updates to be displayed immediately while maintaining a responsive interface without interrupting the reconstruction process. Also kernel settings for riCOM reconstruction and virtual STEM (vSTEM) settings, such as rotation angle due to Lorentz force, kernel size, filter, and the virtual detector size, can be changed during the process without interruption, which is helpful to find suitable settings interactively while spending the lowest amount of dose on the precious sample area.

The riCOM base class is independent of specific camera models and data types, while additional dedicated classes provide live- and file interfaces for given camera types/file formats. This allows for easy extendability of the program, by simply including further interface-classes. The current implementation includes a live- and data interface for the MerlinEM as an example for frame-based data and a filetype interface for the event-based Timepix3 camera and is available on \href{https://github.com/ThFriedrich/riCOM_cpp}{GitHub} under a GPL license.

\subsection*{Materials}
To demonstrate the real time process, an experiment is done on a $SrTiO_3$ focused ion beam (FIB) lamella. Datasets collected from the same material, as well as one recorded in event-based format \citep{jannis_daen_2021_5068510} from a silicalite-1 zeolite sample, are used for testing the effects of different kernel designs and comparing riCOM with other reconstruction methods. The simulated dataset of zeolite is generated by MULTEM \citep{LVAV16_1, LD15_2}, which is a highly optimised multislice simulation package specifically designed for transmission electron microscopy experiments. For the simulation, the thickness of the crystal is chosen to be 20~nm, electron beam energy of 300~keV with convergence angle of 20~mrad, same as the ones used in the experiment. All the datasets and movies recorded on the instrument, including necessary parameters for the reconstruction, can be found in the \href{https://doi.org/10.5281/zenodo.5572123}{online repository} \citep{temporary_repository_address}. 

\subsection*{Equipment}
For real-time reconstruction, the data was acquired with a probe corrected Thermo Fisher Titan$^3$ (X-Ant-TEM), operated at 300~kV with a convergence angle of 20~mrad. The CBEDs are collected with a MerlinEM direct electron detector \citep{ballabriga2011medipix3}.
The zeolite datasets are acquired with Thermo Fisher Themis Z (Advan-TEM) at 200~kV and a custom made Timepix3 detector \citep{poikela2014timepix3} based on an Advapix TP3 camera unit, with a convergence angle of 12~mrad.

\section*{Results and Discussion}
\label{sec:ResDis}

\subsection*{Real Time Reconstruction}

\begin{figure*}[ht]
\centering
\includegraphics[width=0.9\linewidth]{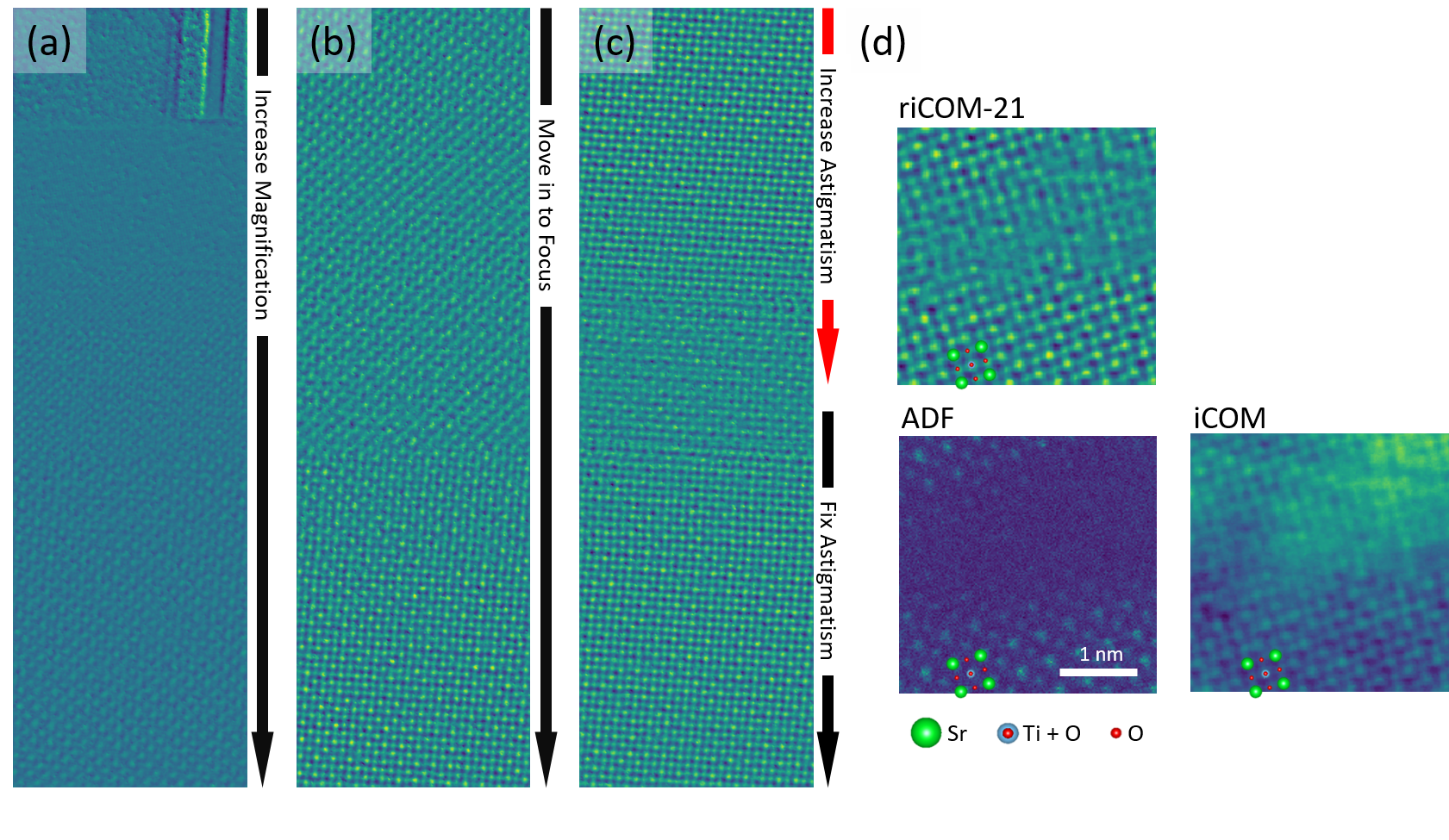}
\caption{Real-time reconstruction of \(\textrm{SrTiO}_\textrm{3}\) while tuning the magnification, defocus, and stigmator. (a) The magnification is increased during the scanning. In the top most part the contrast reveals a layered structure of the FIB lamella, and with increasing magnification the atoms can be captured in the image. (b) Tuning focus is reflected by the change of the shape of atomic columns. (c) Tuning the stigmator affects the electron probe sharpness and also the contrast between atom and vacuum. (d) Simultaneous imaging using riCOM, ADF, and iCOM. RiCOM successfully images the crystalline structure in the centre of the image and the $O$ columns, which is missing in the ADF image. The small kernel size used in riCOM reconstruction reduces long range intensity distribution shown in iCOM.}
\label{fig_live_adjust}
\end{figure*}

To demonstrate riCOM imaging, the software for real time reconstruction is run directly on incoming data during live experiments. The computer receives frames of CBEDs from the detector, and the software reads the data through a TCP socket. Throughout the process, the only extra prior knowledge to be provided to the algorithm is the COM of an undiffracted pattern in vacuum, so that the relative shift of COM at each probe position can be computed. Alternatively, it can also be approximated by averaging the COM from multiple probe positions, thereby omitting any calibration steps, making this method equivalent to more traditional imaging methods regarding ease of use. This step also inherently corrects for systematic shifts of the CBED away from the centre of the detector. While scanning, some of the most basic parameters of the microscopic imaging system are tuned, for example changing the defocus, astigmatism, and magnification, as shown in figure \ref{fig_live_adjust}a-c. The live updated results are recorded in movies included as supporting materials.

Defocus broadens the intensity distribution of the electron probe, and astigmatism has the effect of creating two focal points, making the beam to be first focused in one direction and then the other when travelling down along the optical axis. This would reduce the electron beam sharpness and make the beam elliptical if out of focus, resulting in stretched atom images as can be seen in figure \ref{fig_live_adjust}-b, in the region scanned before achieving right focus.

According to equation \ref{eq:3}, the intensity in the iCOM image equals the cross correlation between the projected electric potential of the material and the probe function, and therefore the reduction in contrast as well as distortions of the atomic features in the riCOM reconstruction is directly related to these beam aberrations. Hence, microscopists can tune optical conditions intuitively to maximise contrast and produce circular atoms with the live updated results.

By changing the magnification during the scanning process the step size is changed accordingly. The live process can still continue, although the intensity needs to be adjusted with a different pixel size $a$ as shown in equation \ref{eq:4}. Besides, the optimal kernel size changes with the magnification, as the spatial frequency of the desired features will be shifted when the step size is changed. However, since the kernel size can be adjusted during the process, a suitable choice can always be found by tuning the kernel size according to the quality of the live updated reconstruction image.

In figure \ref{fig_live_adjust}-d, a riCOM image rendered with a kernel size of 21 is compared to the ADF imaging and iCOM results. Apparent differences can be found in the centre the images, which appears to have a hole according to ADF result but shows some crystalline structure in the riCOM and iCOM images, indicating possible extension of the crystalline material with lower thickness. ADF gives more significant contrast for differences in scattering ability, making it easier to distinguish $Sr$ columns from $Ti + O$ columns, but also reduce intensity of weak scatterers, such as thin regions and the pure $O$ columns, to a level that is completely invisible, while riCOM and iCOM successfully images all three types of columns with a trade-off of less distinction between the columns. On the other hand, atomic structures are blurred by the long range intensity variation in the iCOM result. The origin of this variation could be local strain, misorientation, contamination, charge accumulation, etc., but it is very difficult to pinpoint the actual cause. RiCOM with an appropriate kernel size suppresses these low frequency signals and shows a clear image of atomic columns.

The examples shown in figure \ref{fig_live_adjust} show how riCOM images can be used to fine tune optical systems in a similar manner as using ADF. Moreover, the method is superior to ADF imaging in terms of required electron dose and provides contrast also for the weak scatterers in the object, including thinner regions or atomic columns composed of lighter atoms. The high-pass characteristic of the suitable kernel size has shown to be helpful in highlighting features of higher spatial frequency and reduce low frequency components, but it also means that the contrast interpretation has to be evaluated carefully, especially quantitative analysis, as they can be affected by multiple factors unknowingly.

\begin{figure*}[ht]
\centering
\includegraphics[width=\textwidth]{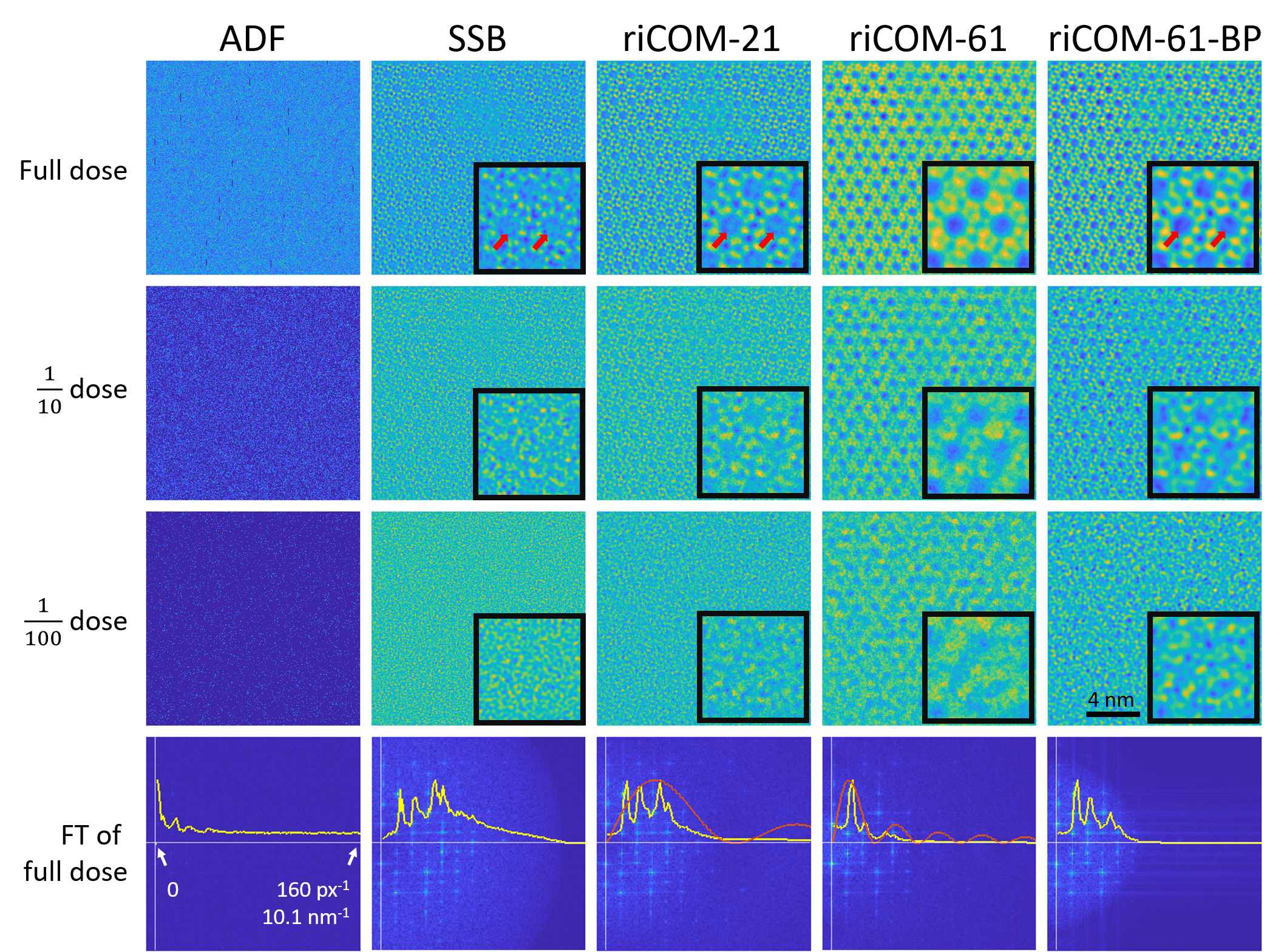}
\caption{
Reconstructed image from an experimental zeolite dataset with different doses (Full dose: 1.27e+4~e/\(\textrm{\AA}^\textrm{2}\)). ADF images are generated by integrating the intensities in the detector area beyond the convergence angle at each probe position. For SSB reconstruction, a frame-based dataset is first generated from the event array, with the detector space binned down to $32\times32$ (8 times smaller). For riCOM reconstruction, three different kernels are used: 21-by-21, 61-by-61, and 61-by-61 with band-pass filter. The effect is however much less significant in other reconstruction method. The insets show magnified versions of the centre of their respective images, and the red arrows point out intensity fluctuations within the holes. The last row shows the Fourier transform of each reconstructed results. The radial averaged frequency spectra presented in yellow curves, and the predicted spectra by line integration approximation in red.}
\label{fig_comp_zeo}
\end{figure*}

\subsection*{Comparison of Reconstruction Methods}

In this section, results from the riCOM reconstruction are compared with other reconstruction methods that have the potential to provide real time imaging. For 4D datasets, ADF images can be computed using a virtual detector integrating all electrons in a specified region of the detector. The summing process is independent of the probe position, and does not require information beyond the scope of a single diffraction pattern, thus making virtual ADF reconstruction possible for real-time visualisation of the dataset. To showcase the performance of riCOM reconstruction, it is compared to both ADF as a traditional imaging mode, and SSB, which is generally considered as a highly dose efficient and quantitative ptychography method. For riCOM reconstruction, three results generated using different kernels are put into comparison, including two kernel sizes and one kernel incorporating a band pass filter.

The dataset used for the comparison is a 4D dataset recorded from a silicalite-1 Zeolite specimen. The dataset is recorded in a sparse array, in which the location where electrons hit the detector and the arrival time is recorded. This type of data format has several advantages over more commonly seen frame-by-frame types at suitable experiment conditions. For instance, in case of low dose imaging, sparse arrays result in datasets many times smaller, since only the detector pixels that successfully capture an electron output data that needs to be recorded, while other inactive pixels remain silent. For riCOM reconstruction, this format also shows its strength in terms of processing speed. Yet another important feature of this format is that the arrival time can be used to adjust the dose in the post reconstruction stage. Since the arrival time of each electron is recorded, the amount of dose put into the reconstruction algorithm can be post-adjusted by reducing the acceptance time from each probe position. For example, with a dataset recorded with a beam dwell time of 6000 ns, if the acceptance time is set to be 2000 ns, any electrons that arrive to the detector after the acceptance time for each probe position will be discarded, and thus effectively reduces the dose to $\frac{1}{3}$ of the original for further processing.   

Accordingly, five data treatment algorithms/setups are used for the experimental data at 3 different dose levels. The results are presented in figure \ref{fig_comp_zeo}. Comparing the images generated by a virtual ADF detector with other reconstruction methods, it is obvious that even with the maximum dose, it is not enough to generate an interpretable ADF image. The vertical lines in the ADF image is a result of camera being inactive for unknown reason, which is discussed in previous work \citep{jannis2021event}. For SSB reconstruction, it includes a process to integrate specific regions in the CBEDs according to their spatial frequency by performing Fourier transformation with respect to probe position. Certain spatial frequencies would gain from larger integration area, and thus creating a band-pass filtering effect \citep{yang2015efficient,o2021contrast}. The riCOM images of smaller kernel size (riCOM-21) are shown to be similar to the SSB results, also manifested by the similarity of their frequency spectra, as low frequency signal is suppressed. For riCOM-101 result, by using a larger kernel size, more components at lower spatial frequencies can be found in the image. These components greatly increases the contrast for the long range structure in the material, such as the pores and framework of the zeolite crystal, but reduces high frequency components, making the short range structures such as atomic columns less clear. This is especially highlighted in the result of \(1/10\) dose. However, by integrating a band pass filter to the big kernel (riCOM-61-BP), noise from the high frequency parts are removed and weights are redistributed to mid-range components from the low frequency end. It results in a much clearer image of the atomic structure even at \(1/10\) dose. The filter used for the last column is designed to remove signals from 3.8~$nm^{-1}$ to 1.14~$nm^{-1}$, with $k_{max} =60~px^{-1}$ and $k_{min} =18~px^{-1}$.

In the third row, only \(1/100\) of the electrons in the dataset is used for imaging. The insufficient amount of electrons introduces a large amount of noise and hides the atomic structure in the images. Yet for the reconstruction result of riCOM-61, the pores within the zeolite framework are preserved in the image. This is possibly due to that features of larger scale are reconstructed from more data points and is thus a result averaged over more possible integration paths. This kind of low frequency components are only supported by kernels of larger size, explaining why other reconstruction methods shown here do not benefit from them and fail to present any meaningful information in the images.

\begin{figure}[ht]
\centering
\includegraphics[width=\linewidth]{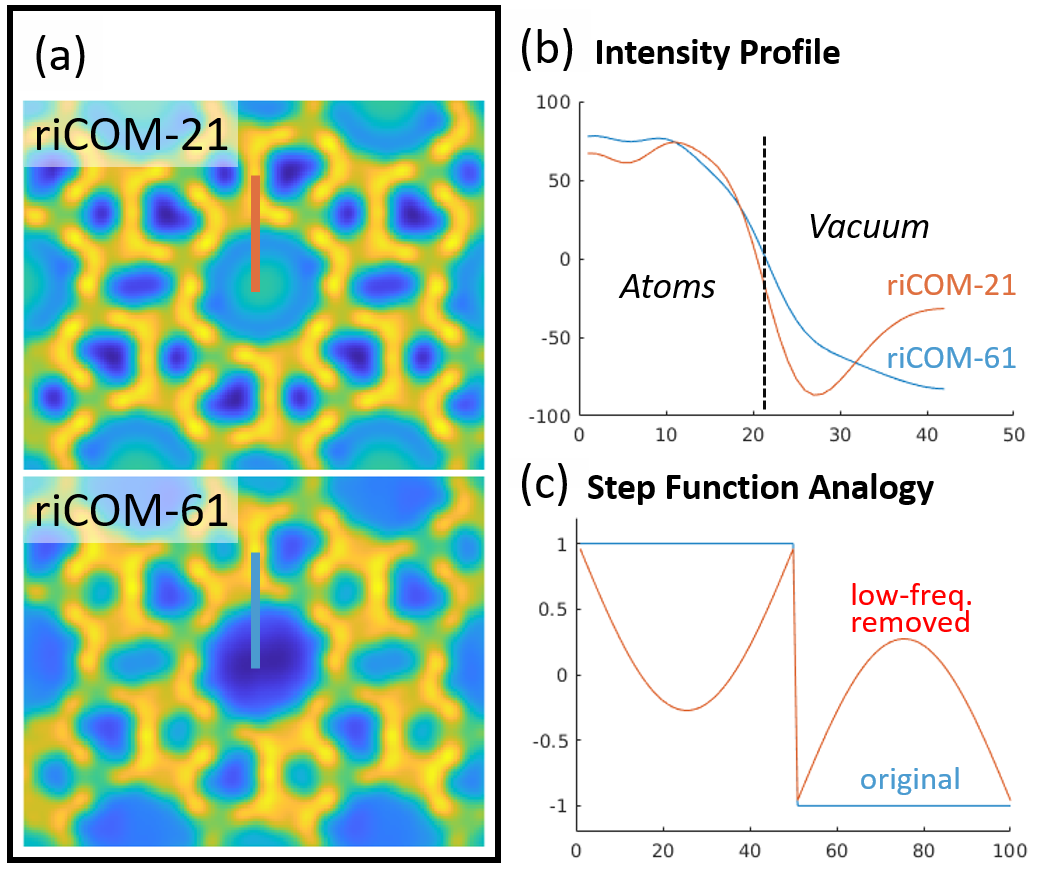}
\caption{
(a) Reconstruction results of a simulated zeolite dataset with different kernel sizes. The red and blue lines indicate the locations of intensity line profiles drawn in subplots (b). (b) The intensity profile shows that the intensity inside the hole area increases in riCOM-21 results but decays in riCOM-61 towards the centre. (c) Step function for analogy shows that removing low frequency components may cause imaging artefacts similar to the ones seen in reconstruction results from smaller kernel sizes.}
\label{fig_comp_zeo2}
\end{figure}

The different reconstruction results also show a disagreement about the content inside of the pores that exist in the zeolite framework. Results from methods that give more weight to the high-frequency components, such as SSB and riCOM-21, show some intensity fluctuation inside of the pores, indicating the possible existence of dopants, yet these do not appear in the riCOM-61 image. In order to understand the reason for that, a 4D dataset was simulated to compare the reconstructed results with different kernel sizes in figure \ref{fig_comp_zeo2}. An intensity profile is drawn over the atom framework into the pore (Fig. \ref{fig_comp_zeo2}-a), which is indeed vacuum as designed for the simulation. The profile reveals that for riCOM-21, the intensity increases, while riCOM-61 shows a monotonic decay towards the centre of the pore (Fig. \ref{fig_comp_zeo2}-b). The intensity increase for riCOM-21 does not originate in the projected atomic potential, since it can only decay when moving further away from the atoms. 

To investigate the origin of this false intensity, the Fourier transformed riCOM images are analysed (Fig. \ref{fig_comp_zeo}), The bright spots at the lowest frequency correspond to the periodic structure of the pores and framework. The intensity of these spots are greatly reduced in riCOM-21 but supported in riCOM-61, indicated by the approximated weighting function based on the kernel size (red curves, equation given in \ref{eq:freq_approx}). This causes major differences to features that necessarily rely on such low frequency signals. To illustrate the principle, we simplify the atom framework and the pore using a step function (Fig. \ref{fig_comp_zeo2}-c). By removing the low frequency components, the step becomes a curve with a concave and a convex segment in the regions of the high and the low step, respectively. This step function analogy conceptually captures the differences between the zeolite framework and the holes and explains the protruding intensity in the hole for riCOM-21 as the effect of reduced low-frequency components. For riCOM-61 such components are included by the larger kernel size, so that no such phantom intensity can be found in the same area.

These examples show that the proposed method, like many other reconstruction methods, is capable of providing extra information compared to traditional imaging methods. RiCOM also shows great dose efficiency, allowing high quality reconstruction results under low dose conditions. The freedom to use different kernel sizes grants users the ability to tune the desired spatial frequency range, which is very important in order to avoid misinterpretation of details in the image. Including more low-frequency components, has shown to enable the reconstruction of long range structures of the object with even lower amounts of electrons. This could be very useful for microscope operators for imaging objects of larger scale.

\section*{Conclusion}
In this paper we propose and demonstrate a reconstruction method for real time STEM based on the integrated centre of mass that is applicable to any kind of segmented detector dataset, including but not limited to 4D-STEM. Through derivation of the physical formulation, we illustrate the physical relevance and the benefits for numerically efficient implementations of this approach, motivating the application particularly in real-time imaging scenarios. The freedom to change the size of the kernel or incorporating filters are also discussed, with examples showing their effect.  

It is shown that riCOM can effectively reproduce iCOM results, but allows for more flexibility in terms of selecting contributing spatial frequencies. The method, including frequency band pass filtering depends only on the individual intensity distribution (or CBED) at its corresponding real space location, which in combination with a rather simple algorithm, creates a uniquely flexible and fast reconstruction method that requires very little user input. We further present a well optimised, interactive GUI implementation, developed in standard C++ and published open source on \href{https://github.com/ThFriedrich/riCOM_cpp}{GitHub}. 

Demonstrations of the method on an operating microscope shows that firstly, the process is fast enough to keep up with the highest frame rate supported by currently available detectors, and secondly, providing dynamic feedback to the microscope operator when tuning and optimising the microscope parameters. This ability enables swift search of the sample, or region of interest, as well as adjustments of the imaging conditions, at potentially very low dose conditions. The algorithm can run on any kind of data from which the centre of mass of the electron diffraction pattern, or derivatives of COM such as DPC signals, can be calculated, and therefore it is by no means limited to the hardware demonstrated in this paper.

Comparisons with results of other non-iterative reconstruction methods show that riCOM renders high quality images on par with established methods, even at very low doses. The pros and cons of using different frequency components are discussed. Users can accordingly choose the most suitable designs of kernels, and run simultaneously other imaging forming methods, in order to reach the highest dose efficiency or extract the most amount of knowledge from the investigated sample in real time.
\\

\noindent\small\color{Maroon}\textbf{Acknowledgements }\color{Black}
We acknowledge funding from the European Research Council (ERC) under the European Union’s Horizon 2020 research and innovation program (Grant Agreement No. 770887 PICOMETRICS) and funding from the European Union’s Horizon 2020 research and innovation program under grant agreement No. 823717 ESTEEM3. J.V. and S.V.A acknowledge funding from the University of Antwerp through a TOP BOF project. The direct electron detector (Merlin, Medipix3, Quantum Detectors) was funded by the Hercules fund from the Flemish Government. 
Further, the authors gratefully acknowledge Xiaobin Xie (Sichuan University, P. R. China) for providing the STO sample used in this study.

\normalsize

%\bibliographystyle{MandM}
%\bibliography{ref}

\newpage
\onecolumn
\appendix
\renewcommand\thefigure{S\arabic{figure}} 
\setcounter{figure}{0}   

\section*{Supporting Information}

\subsection*{Code and Data}

Datsets used in this study and video recordings of live imaging session using riCOM can be found online at the open data platform Zenodo: 
\begin{tabbing}
authors: \=  Chu-Ping Yu and Thomas Friedrich \\
title: \> Real Time Integration Center of Mass (riCOM) Reconstruction for 4D-STEM \\
doi: \> 10.5281/zenodo.5572123 \\
url: \> \href{https://doi.org/10.5281/zenodo.5572123}{https://doi.org/10.5281/zenodo.5572123}\\
\end{tabbing}

\noindent The Source Code and precompiled binaries for Windows 10 and Ubuntu OS can be found on GitHub:
\begin{tabbing}
authors: \=  Thomas Friedrich and Chu-Ping Yu \\
title: \> riCOM\textunderscore cpp \\
license: \> GNU GPL3 \\
url: \> \href{https://github.com/ThFriedrich/riCOM_cpp}{https://github.com/ThFriedrich/riCOM\textunderscore cpp} 
\end{tabbing}

\subsection*{Derivations}

To prove equation 2 holds, we demonstrate moving the gradient out of the cross correlation with this example. Here we assume three functions, a, b, and c, with the relationship

\begin{align*}
a(x,y) &= b(x,y) \star c(x,y).
\end{align*}

By taking Fourier transform on both side we have

\begin{align*}
A(\hat{x},\hat{y}) &= \bar{B}(\hat{x},\hat{y}) \times C(\hat{x},\hat{y}),
\end{align*}

with which the equation can be rewritten with function a and the inverse Fourier transformed B and C

\begin{align*}
a(x,y) &= F^{-1}\{ \bar{B}(\hat{x},\hat{y}) \times C(\hat{x},\hat{y}) \} \\
&= \int{ \int{ \bar{B}(\hat{x},\hat{y}) C(\hat{x},\hat{y}) e^{ix\hat{x}} e^{iy\hat{y}} d\hat{x} d\hat{y} } }.
\end{align*}

Taking partial derivative of x from both side,

\begin{align*}
\frac{\partial a(x,y)}{\partial x} &= \frac{\partial}{\partial x} F^{-1}\{ \bar{B}(\hat{x},\hat{y}) \times C(\hat{x},\hat{y}) \} \\
&= \int{ \int{ \bar{B}(\hat{x},\hat{y}) C(\hat{x},\hat{y}) \frac{\partial e^{ix\hat{x}}}{\partial x} e^{iy\hat{y}} d\hat{x} d\hat{y} } } \\
&= i\hat{x} \int{ \int{ \bar{B}(\hat{x},\hat{y}) C(\hat{x},\hat{y}) e^{ix\hat{x}} e^{iy\hat{y}} d\hat{x} d\hat{y} } }.
\end{align*}

By putting \(i\hat{x}\) together with function C, and make use of the derivative rule of fourier transform, we have

\begin{align*}
\int{ \int{ \bar{B}(\hat{x},\hat{y}) (i\hat{x} C(\hat{x},\hat{y}))  e^{ix\hat{x}} e^{iy\hat{y}} d\hat{x} d\hat{y} } } &= \int{ \int{ \bar{B}(\hat{x},\hat{y}) F\{ \frac{\partial c(x,y)}{\partial x} \}  e^{ix\hat{x}} e^{iy\hat{y}} d\hat{x} d\hat{y} } } \\
\frac{\partial}{\partial x} F^{-1}\{ \bar{B}(\hat{x},\hat{y}) \times C(\hat{x},\hat{y}) \} &= F^{-1}\{ \bar{B}(\hat{x},\hat{y}) \times F\{ \frac{\partial c(\hat{x},\hat{y})}{\partial x} \} \} \\
\frac{\partial}{\partial x} ( b \star c ) &= b \star \frac{\partial c}{\partial x} 
\end{align*}

The gradient is represented as \( \nabla  = \frac{\partial}{\partial x} \cdot \boldsymbol{n_x} + \frac{\partial}{\partial y} \cdot \boldsymbol{n_y} \), where \(\boldsymbol{n_x}\) and \(\boldsymbol{n_y}\) are the unit vectors in both directions. This shows that we simply need to take another partial derivative with y to acquire the gradient of the function \( b \star c \), and due to the associative property of cross correlation, this applies to the right side of the equation as well. So in the end we have   

\begin{align*}
\nabla( b \star c ) = b \star \nabla c.
\end{align*}

\subsection*{Additional Figures}

\begin{figure}[ht]
\centering
\includegraphics[width=0.7\linewidth]{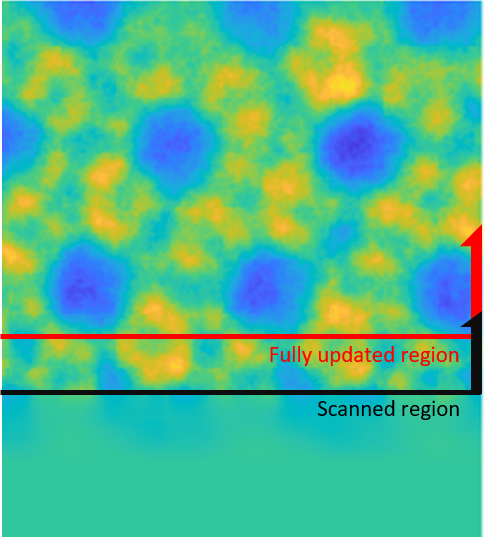}
\caption{
A halfway reconstructed image. Scanning leads the full update of iCOM image by half of the kernel size, but the result shows that atomic features are already visible before the image is fully updated.}
\end{figure}

\end{document}